\documentclass [twocolumn,aps,pra,noshowpacs]{revtex4-1}
\usepackage{epsfig,graphicx,amssymb,color,amssymb,amsmath,mathrsfs,bm}
\pdfoutput=1

\begin{document}

\title{Interplay of gain and loss in arrays of nonlinear plasmonic nanoparticles: toward parametric downconversion and amplification}

\author{Syed A. Shah}
\affiliation{Center for Nonlinear Studies (CNLS), Theoretical Division, Los Alamos National Laboratory, Los Alamos, NM 87545, USA}

\author{Michael R. Clark}
\affiliation{Center for Nonlinear Studies (CNLS), Theoretical Division, Los Alamos National Laboratory, Los Alamos, NM 87545, USA}
\affiliation{Department of Physics, Arizona State University, Tempe, AZ 85287, USA}

\author{Joseph Zyss}
\affiliation{LUMIN Laboratory and Institut d’Alembert, Ecole Normale Supérieure Paris Saclay, CNRS, Université Paris-Saclay, 4, Avenue des Sciences, Gif-sur-Yvette, France}

\author{Maxim Sukharev}
\affiliation{Department of Physics, Arizona State University, Tempe, AZ 85287, USA}
\affiliation{College of Integrative Sciences and Arts, Arizona State University, Mesa, Arizona 85212, USA}

\author{Andrei Piryatinski*}
\affiliation{Center for Nonlinear Studies (CNLS), Theoretical Division, Los Alamos National Laboratory, Los Alamos, NM 87545, USA}

\begin{abstract}
With the help of a theoretical model and finite-difference-time-domain simulations based on the hydrodynamic-Maxwell model, we examine the effect of difference-frequency generation in an array of L-shaped metal nano-particles characterized by intrinsic plasmonic nonlinearity. The outcomes of the calculations reveal the spectral interplay of the gain and loss in the vicinity of the fundamental frequency of the localized surface-plasmon resonances. Subsequently, we identify different array thicknesses and pumping regimes facilitating parametric amplification and spontaneous parametric down-conversion.  Our results suggest that the parametric amplification regime becomes feasible on a scale of hundreds of nanometers and spontaneous parametric downconversion on the scale of tens of nanometers, opening up new exciting opportunities for developing building blocks of photonic metasurfaces.
\end{abstract}

\date{\today}
 
\maketitle


\section{Introduction}
Nonlinear nano-optics has been a subject of increasing interest in the past few decades. With significant advancements in experimental capabilities enabling the concentration of light into sub-diffraction volumes, numerous nonlinear optical phenomena, once predominantly studied in macroscopic systems, are now being explored at the nanoscale \cite{kauranen2012nonlinear}. A wide variety of metal nanoparticles (MNPs) shapes and arrays of nanoholes have been employed to investigate phenomena such as second harmonic generation~\cite{butet2015optical,Ellenbogen:2017}, four-wave mixing~\cite{almeida2015rational}, and many other nonlinear processes~\cite{panoiu2018nonlinear}. For instance, the second harmonic generation was demonstrated to be a very sensitive optical tool for detecting spatial variations in MNPs shapes at the nanoscale level \cite{maekawa2020wavelength}. 

While significant attention has been paid to the harmonic generation using plasmonic nanostructures~\cite{panoiu2018nonlinear}, reports on the reverse process of difference frequency generation (DFG) have been limited. Within this context, the intrinsic nonlinear response of MNP arrays has been explored theoretically and experimentally demonstrated to support an efficient generation of THz radiation via the DFG~\cite{Huttunen_PRA:2018,Sideris:19}. A design of a V-shaped double resonant nanoantenna has been proposed to leverage the intrinsic coupling between plasmon modes enhancing the DFG efficiency~\cite{deLucaJOSAB:2019}. Additionally, tailored surface-plasmon resonances at the pump, signal, and idler frequencies of the Au-SiO$_2$-shell have been taken advantage of to enhance extrinsic nonlinearity of the BaTiO$_3$ core facilitating the DFG process towards achieving parametric amplification~\cite{zhangNanoLet:2016}. Moreover, the use of metal surfaces has been explored theoretically and experimentally to enhance the extrinsic nonlinearity of a substrate for spontaneous parametric down-conversion (SPDC) purposes~\cite{Loot_and_Hizhnyakov,Hizhnyakov_exp}. {\color{black} These findings are particularly significant for the development of probabilistic quantum photon sources as components of quantum metasurfaces.~\cite{kivshar:2021}}

In the DFG studies considering the intrinsic nonlinearity mechanisms, theoretical efforts were directed towards developing and applying perturbative hydrodynamic models capable of accurate description of the plasmonic modes, and engineering the intermode couplings~\cite{Sideris:19,deLucaJOSAB:2019}. However, these analyses have not addressed an important issue of the interplay between plasmon gain and loss, crucial for achieving parametric amplification and controlling the down-conversion processes. In this letter, we address this issue in the case of L-shaped (a limiting configuration of the V-shaped) MNP arrays. Leveraging the fact that in such MNPs, both linear and nonlinear responses are dominated by dipolar contributions~\cite{panoiu2018nonlinear}, we introduce a minimal theoretical model for DFG and parametric amplification. This model provides insights into the spectral properties of the parametric gain and loss, enabling us to identify the length scale supporting efficient parametric amplification. Subsequently, we present and interpret the results of our  finite-difference time-domain (FDTD) simulations, employing the semiclassical hydrodynamic model~\cite{Scalora_Vincenti_de_Ceglia_Roppo_Centini_Akozbek_Bloemer_2010} to characterize the surface-plasmon response of the electron gas in the MNPs. {\color{black} In the end, we employ a semiclassical approach based on our minimal model to examine the SPDC yield of the MNP array.}   

\begin{figure}[t]
\centering
\includegraphics[width=0.45\textwidth]{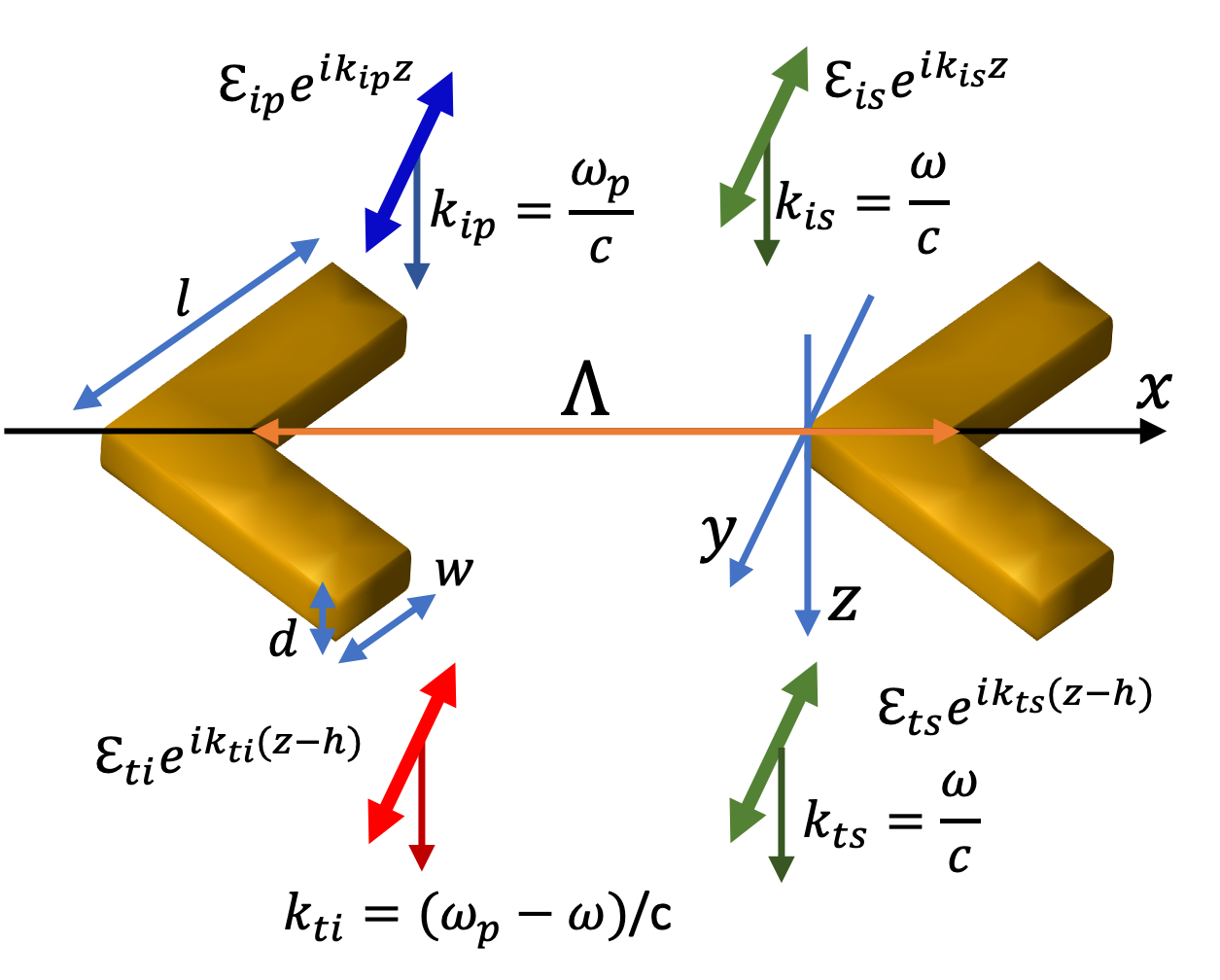}
\caption{
Schematic of a 1D array of a period $\Lambda=400$~nm, with repeat unit formed by an L-shaped MNP characterized by dimensions $d=30$ nm, $w=40$ nm, and $l=130$~nm. Correspondingly, ${\cal E}_{ip}e^{ik_{ip}z}$ and ${\cal E}_{is}e^{ik_{is}z}$ represent the normally incident pump and signal electric fields, while ${\cal E}_{ti}e^{ik_{ti}(z-h)}$ and ${\cal E}_{ts}e^{ik_{ts}(z-h)}$ represent the transmitted idler and signal electric fields. The transmitted pump field is not shown in the plot.
}
\label{fig:1}
\end{figure}
\section{Theoretical Model} 
As depicted in Fig.~\ref{fig:1}, we examine a 1D array of L-shaped metal nanoparticles (MNPs) aligned in the $x$-direction with a period $\Lambda$ within the $(x,y)$-plane. The array has thickness $d$ in the $z$-direction with $z=0$ attributed to the top boundary. This array is immersed in a dielectric medium with a refractive index $n_m=1$ and subjected to monochromatic light of frequency $\omega$  incident perpendicular to the array plane, where the electric field is linearly polarized in the $y$-direction. Accordingly, we describe the array as an optical medium with the refractive index, $n(\omega) =\sqrt{1 + \chi^{(1)}(\omega)}$, ~a function of the array's linear susceptibility $\chi^{(1)}(\omega)$. In general, the linear susceptibility of a periodic array of MNPs contains both local surface plasmon resonances and collective surface lattice resonances~\cite{deAbajo_RMP:2007,Cherqui_Bourgeois_Wang_Schatz_2019}. In our calculations, we focus on the spectral region containing the local resonance associated with the $y$-polarized eigenmode of the L-shaped MNP~\cite{Sukharev_PRB_2007}. For this eigenmode, we approximate the linear susceptibility of the array by the Lorentzian function 
\begin{eqnarray}
\label{chi-mnp}
\chi^{(1)}(\omega) =\frac{p_\texttt{sp}^2}{v\varepsilon_o\hbar\left(\omega_\texttt{sp}-\omega -i\gamma_\texttt{sp}\right)},
\end{eqnarray}
depending on the resonance energy $\hbar\omega_\texttt{sp}$, associated linewidth (damping) $\hbar\gamma_{sp}$, transition dipole $p_\texttt{sp}$, repeat unit volume $v$, and the vacuum permittivity $\varepsilon_o$.

{\color{black} The DFG involves a three-wave mixing process where the pump of frequency $\omega_p$, propagating in the $z$-direction inside the layer of MNPs, transfers its energy to the signal and idler fields of frequencies $\omega$ and $\omega-\omega_p$, respectively. As we derive in Sec.~1 of Supplement~1, for finite incident signal intensity $I_{is}$, and zero incident idler field, the transmitted ($z>d$) normalized electric field intensity of the signal and idler have functional forms 
\begin{eqnarray}
\label{Its}
\hspace{-0.6cm}I_{ts} (\omega)/I_{is}&=& \left|\frac{4 n(\omega)}{[1+n(\omega)]^2}\right|^2\left|\cosh\left[\xi(\omega) d\right]
\right.\\\nonumber&-&\left.
 \frac{i \Delta k(\omega)}{2\xi(\omega)}\sinh\left[\xi(\omega) d\right]\right|^2 e^{\textrm{Im}\Delta k(\omega) d},
\\\label{Iti}
\hspace{-0.6cm}I_{ti} (\omega_p -\omega)/I_{is}&=& 
    \left|\frac{4n(\omega_p -\omega)g(\omega_p-\omega)}{\left[1+n^*(\omega)\right]\left[1+n(\omega_p -\omega)\right]\xi(\omega)}\right|^2
\\\nonumber&\times&
\left|\sinh\left[\xi(\omega) d\right]\right|^2~e^{\textrm{Im}\Delta k(\omega)d},
\end{eqnarray}
respectively.} Here, the shorthand notation $\xi(\omega)$ stays for the complex parameter
\begin{eqnarray}
\label{xi-def}
\xi(\omega) = \sqrt{g(\omega)g^{*}(\omega_p-\omega)-\Delta k^2(\omega)/4},
\end{eqnarray}
defining the gain $G(\omega)=\text{Re}[\xi(\omega)]$ competing with the loss $\Gamma(\omega)=\text{Im}[\Delta k(\omega) /2]$. {\color{black} The signal-idler coupling parameter is
\begin{eqnarray}
\label{gs-def}
 g(\omega) =\left(\frac{2I_p}{\varepsilon_o c}\right)^{1/2} \frac{2i\omega\chi^{(2)}(\omega_p;\omega;\omega_p-\omega)}
        {cn(\omega)\left[1+n(\omega_p)\right]},
\end{eqnarray}
and contains the second-order nonlinear susceptibility which according to the anharmonic oscillator model, we represent  as
\begin{equation}
\centering
\label{chi2-mnp}
\chi^{(2)}(\omega_p;\omega,\omega_p-\omega) ={\cal A}\chi^{(1)}(\omega_p)\chi^{(1)}(\omega)\chi^{(1)}(\omega_p-\omega),
\end{equation}
where $\cal A$ is the anharmonicity constant. Finally, the wave vector mismatch entering Eqs.~(\ref{Its}) -- (\ref{xi-def})  is
\begin{equation}
\centering
\label{dk-def}
\Delta k (\omega) = k(\omega_p) -k(\omega)-k^*({\omega_p-\omega}).
\end{equation}
Since, the pump is set within a lossless frequency range, $k(\omega_p)=n(\omega_p)\omega_p/c$ is real function. In contrast, the wave vectors of the signal, $k(\omega)=n(\omega)\omega/c$, and idler, $k(\omega_p - \omega)=n(\omega_p-\omega)(\omega_p-\omega)/c$, are complex, since, we tune them within proximity of the lossy plasmonic resonance.}

\begin{figure}[t]
\centering
   \includegraphics[width=0.95\linewidth]{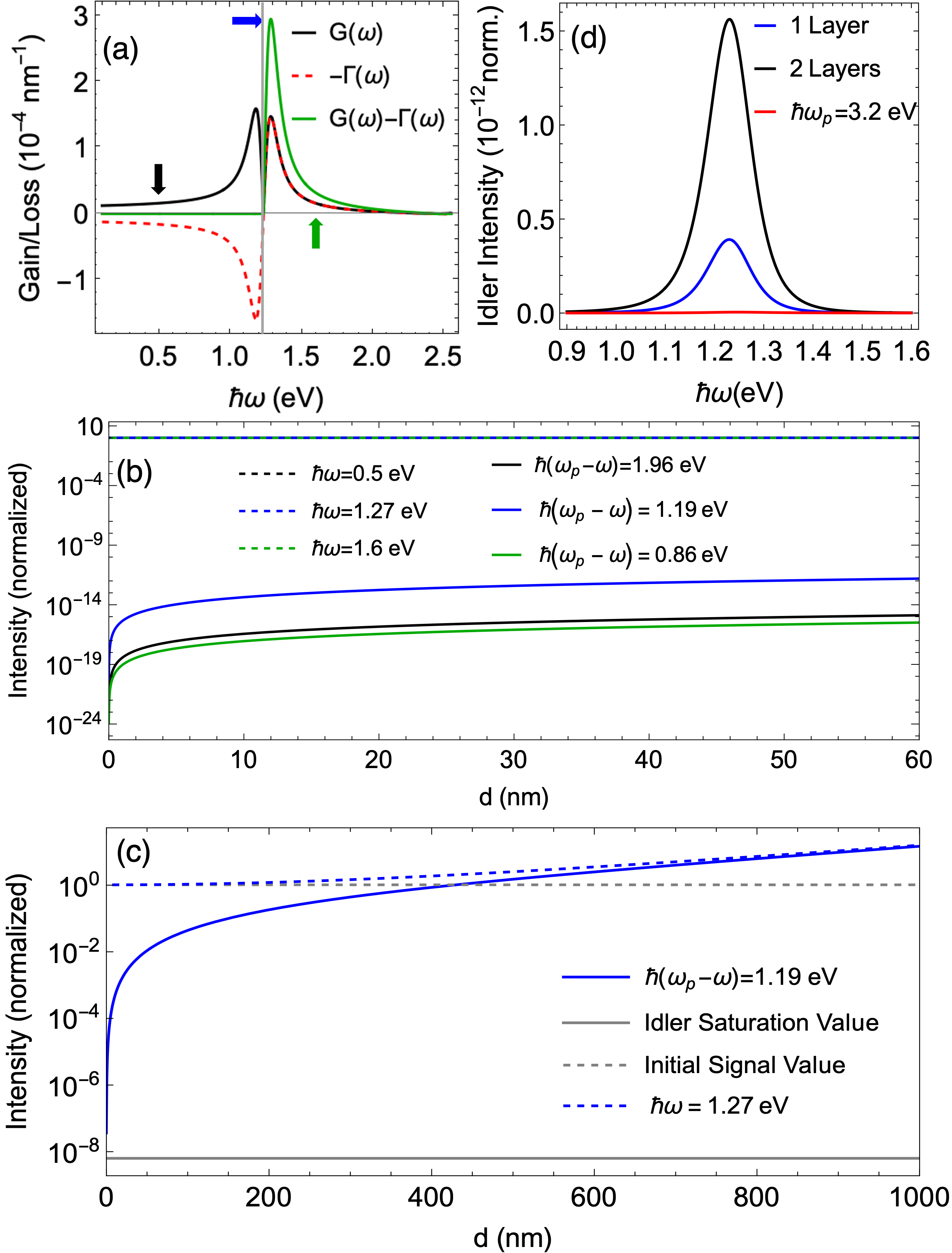}
   \caption{(a) The spectral behavior of the gain $G(\omega)$, loss $\Gamma(\omega)$, and their net value $G(\omega)-\Gamma(\omega)$ in the vicinity of the surface plasmon resonance, marked by the vertical gray line at $\hbar\omega=1.232$~eV.
(b) Calculated in the weak coupling regime, normalized intensity of the transmitted idler (signal), $I_{ti}/I_{is}$ ($I_{ts}/I_{is}$), shown by solid (dashed) curves as a function of the array thickness $d$. Three signal energies and their associated gain/loss values, marked with arrows in panel~(a), are used in the calculations. The pump energy is set to $\hbar\omega_p=2.464$~eV.
(c) Normalized intensities of the signal and idler, calculated for the same parameters as in (b) but with the signal-idler coupling increased by a factor of $10^5$. For comparison, the idler saturation value attributed to the weak coupling regime and the initial signal value are presented by the gray lines. (d) Normalized idler intensities as a function of signal frequency for a single  layer ($d=30$ nm) in blue and double layer ($d=60$ nm) in black. Additionally, red curve indicates the single layer idler for $\omega_p$ offset from  $2\omega_{sp}$ }
   \label{fig:2} 
\end{figure}

To parameterize our model (see Sec.~2 in Supplement~1), we centered our attention on the Au MNPs of dimensions specified in Fig.~\ref{fig:1} and {\color{black} performed FDTD simulations. This yielded the plasmon energy $\hbar \omega_{sp} = 1.232$ eV, linewidth $\hbar \gamma_{sp}=0.077 $~eV, transition dipole $p_{sp}= 4.75\times 10^{3}$ D, and the anharmonicity parameter ${\cal A}=4.0$~pm/V. Finally setting the incident pump field to ${\cal E}_{ip} = 1.25\times 10^8~\text{V/m}$ and} frequency to the second harmonic of the plasmon resonance $\omega_p=2\omega_{sp}$, we employed Eqs.~(\ref{Its})--(\ref{dk-def}) to calculate normalized transmitted idler (signal) intensity, $I_{ti}/I_{is}$ ($I_{ts}/I_{is}$) as a function of the array thickness $d$ and depending on their behaviour, examined the interplay between parametric gain $G(\omega)$ and loss $\Gamma(\omega)$.  

The results are presented in Fig.~\ref{fig:2}. Panel~(a) shows that below the surface plasmon resonance (vertical gray line), the loss (red dashed curve) is compensated by the gain (black solid curve), resulting in a net value of zero (solid green curve). Above the resonance, the loss parameter changes sign, causing the gain to double as observed in the green curve rising. This behavior can be attributed to the fact that the contribution from the signal-idler coupling is significantly smaller than the wave vector mismatch, i.e., $|g(\omega)g^{*}(\omega_p-\omega)|\ll|\Delta k^2(\omega)/4|$. According to Eq.~(\ref{xi-def}), in such weak coupling regime, the ratio $\xi(\omega)\approx\pm i\Delta k(\omega)$ and particularly $\Gamma(\omega)\approx\pm G(\omega)$. The latter equality precisely shows that depending on the sign of the loss term, it either cancels or doubles the gain as observed in the plot.  

In Fig.~\ref{fig:2}(b), we present normalized intensities of the transmitted signal (dashed curves) and transmitted idler (solid curves)  as a function of the array thickness $d$ for three distinct signal energies indicated by the arrows in panel~(a). These energies correspond to spectral regions below the surface plasmon resonance (black curve), at the peak of the gain (blue curve), and on the blue tail of the gain (green curve). We observe a fast rise in the idler intensity occurring on the scale of tens of nanometers. As anticipated, an increase in the gain value leads to a corresponding enhancement in the idler amplitude for a fixed $d$. 

In the weak coupling limit, all idler curves tend to approach corresponding saturation plateaus defined by $\left|\frac{4n(\omega_p -\omega)g(\omega_p-\omega)}{\left[1+n^*(\omega)\right]\left[1+n(\omega_p -\omega)\right]\xi(\omega)}\right|^2$. However, the signal in Eq.~(\ref{Its}) becomes independent of the gain and loss parameter, maintaining its initial value of $\left|\frac{4n(\omega)}{[1+n(\omega)]^2}\right|^2$ as represented by the dashed lines in Fig.~\ref{fig:2}(b). {\color{black} The saturation values are sensitive to the refractive index of the embedding material, initially set to $n_m=1$ and subsequently examined with $n_m=1.5$ (see Sec.~3 in Supplement~1). The calculations reveal that an increase in refractive index lowers the signal and idler saturation values by two orders of magnitude and the gain values about fourfold. Additionally (see Sec.4 in Supplement~1), we evaluate the power-conversion efficiency, measured in the experiment, for all curves in Fig.~\ref{fig:2}(b), yielding saturation values ranging between $10^{-13}-10^{-9}$ \%/W.} 

While the signal-idler coupling  is notably small in comparison to the momentum mismatch, its effect remains non-negligible. It contributes to the nonzero value of the idler ensuring the occurrence of downconversion. Moreover, it introduces a subtle imbalance in the equilibrium between the gain and loss parameters, ultimately leading to the exponential growth of both the signal and idler beyond the saturation plateau. Based on our estimate for the given parameters, such deviations show up at $d\gtrsim 10^5$nm, indicating the parametric amplification regime. To reach this regime, the use of an optical cavity would be necessary. Conversely, an order of $10^5$ increase in the coupling parameter allowed us to observe the amplification regime on the length scale of hundreds of nanometers at the peak gain. This is demonstrated in panel~(c) of Fig.~\ref{fig:2}. The enhancement can be realized by employing MNPs with a high anharmonicity constant ${\cal A}$ and/or by increasing the pump intensity $|I_{ip}|^2$. Finally, Fig.~\ref{fig:2}~(d) shows normalized idler intensity from a single layer ($d=30$ nm) as well as a double layer (modeled with $d=60$ nm) as a function of signal frequency with pump fixed at $2\omega_{sp}$. Additionally, a single layer idler with off-resonance pump at $\hbar \omega_p = 3.2$ eV (shown in red) demonstrates a stark reduction in downconversion efficiency in this spectral range.

\section{Numerical Results} 
Now, we proceed to demonstrate feasibility of the DFG via the FDTD numerical simulations. We treat the linear and nonlinear plasmonic response of Au MNPs on the same footing using the hydrodynamic model for the conduction electrons coupled with the Maxwell equations for the electromagnetic field~\cite{Scalora_Vincenti_de_Ceglia_Roppo_Centini_Akozbek_Bloemer_2010, Sukharev_Drobnyh_Pachter_2022}.
The hydrodynamic model was supplied with the following set of input parameters for Au: electron number density $n_0=5.9\times10^{28}\mathrm{m}^{-3}$, effective mass of electron $m_e^*=1.66~m_e$ where $m_e$ is electron mass in vacuum, the damping parameter $\hbar\gamma_e = 0.181$~eV, and the Fermi energy  $E_F = 5.53$ eV. The geometrical parameters and the incident optical pulse configuration remain unchanged, as depicted in Fig.~\ref{fig:1}. Power spectra are obtained at normal incidence by applying 500 fs long pump with an amplitude of 5$\times$10$^7$ V/m and a low intense signal of the same duration with an amplitude 5$\times$10$^4$ V/m. Polarization of both the pump and the signal is set along one of the arms of MNP. Additionally, we investigate a two-layer array with the layer separation of 150 nm, effectively doubling the optical path in the $z$-direction. 

\begin{figure}[t]
     \centering
      \includegraphics[width=0.95\linewidth]{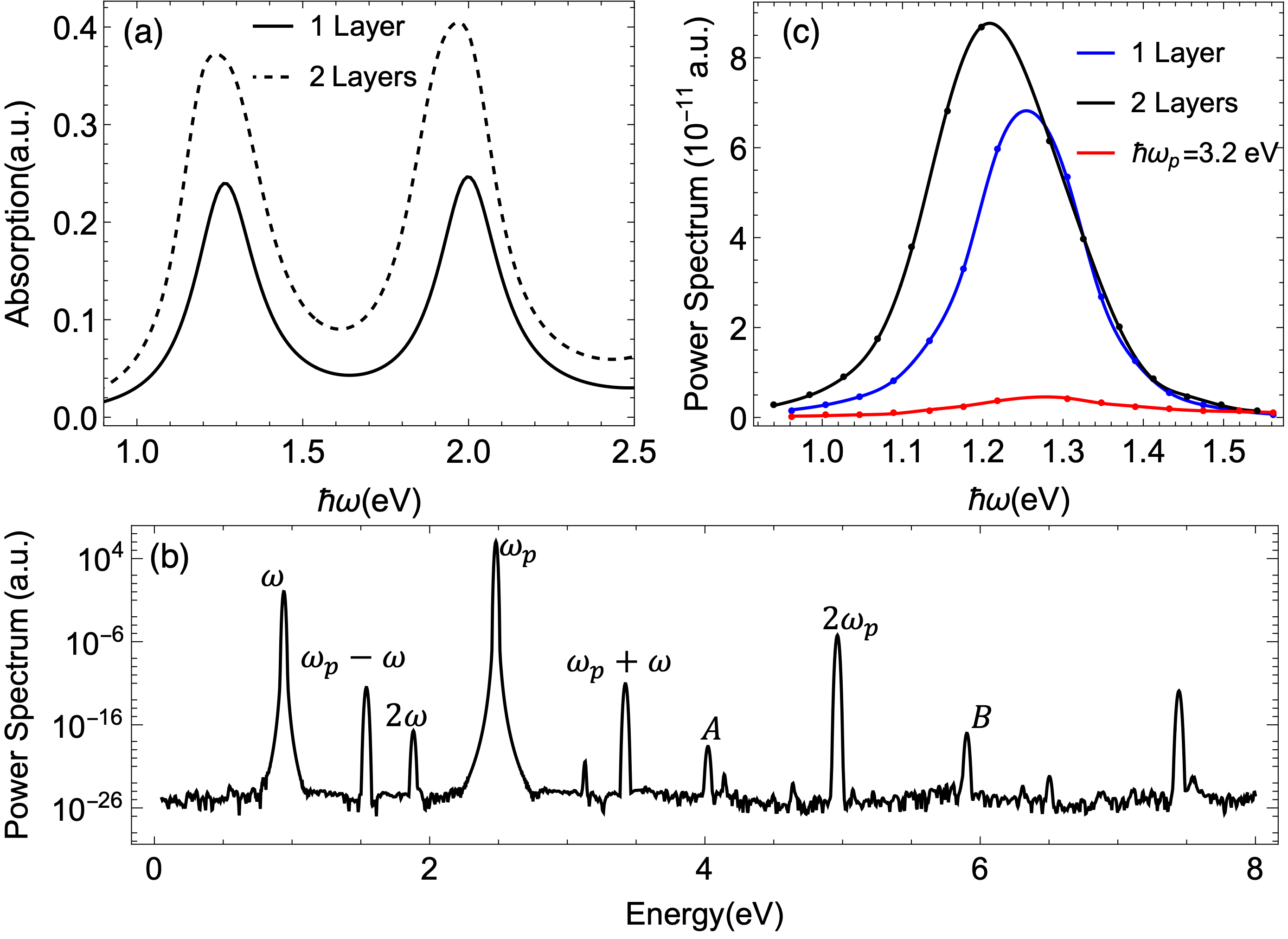}
        \caption{The FDTD simulation results for single- and double-layer arrays of L-shaped Au MNPs: (a) Linear absorption spectra. (b) The power spectra from the double layer array containing, linear response subject to simultaneous pump $\omega_p$ and signal $\omega$ excitation and their three- and four-wave mixing resonance. (c) Comparison of the single- and double-array idler intensities as a function of the signal frequency $\omega$. For the black and blue (red) cures the pump energy is twice the $y$-polarized plasmon resonance frequency, $\hbar\omega_p=2.464$~eV. The red curve shows the single-layer transmitted idler intensity associated with the pump energy offset from twice of the surface plasmon resonance. All idler curves are normalized similar to those in Fig.~\ref{fig:2}~(d).}
        \label{fig:3}
\end{figure}

The results of the numerical simulations are presented in Fig.~\ref{fig:3}. In panel (a), the linear absorption spectra show two resonances. The first, located at $1.232$ eV, as discussed above, is associated with the $y$-polarized local surface plasmon mode. The second resonance at $2.00$~eV is attributed to the $x$-polarized eigenmode~\cite{Sukharev_PRB_2007}. A comparison of the linear absorption between the single-layer (solid curve) and double-layer (dashed curve) configurations demonstrates that the interlayer plasmonic coupling is sufficiently weak and does not have significant effect on the resonance positions. The hydrodynamic-Maxwell model we employ inherently incorporates all orders of the plasmonic response. The calculated power spectrum presented in panel~(b) shows a range of three-wave mixing processes. These include second harmonic generations due to both the pump ($\omega_p$) and the signal ($\omega$), sum-frequency generation between the pump and signal, and DFG yielding the idler at $\omega_p-\omega$. The spectrum also exhibits the features labeled as $A$ and $B$ corresponding to the four-wave mixing processes $2\omega_p-\omega$ and $2\omega_p+\omega$, respectively. Below, we focus on the DFG process only.

Figure~\ref{fig:3}~(c) demonstrates the variation of idler intensity as a function of the signal frequency for single (blue curve) and double (black curve) layers with pump fixed at $\omega_p=2\omega_{sp}$. In conjunction with our minimal model, the idler intensity peaks around the surface plasmon resonance with double layers showing increased idler production. Additionally, the red curve shows the idler intensity from a single layer with the pump detuned from the second harmonic of the plasmon resonance to $\hbar \omega_p=3.2$~eV, exhibiting a pronounced decrease in the downconversion efficiency as the minimal model predicts. Furthermore, for all three curves, the corresponding signal intensity (not shown in the plot) do not change while propagating through the array in $z$-direction.  In accordance with analytical findings of Fig.~\ref{fig:2}~(b), we conclude that the DFG process takes place in the weak coupling regime.

\section{SPDC Yield}

\begin{figure}[t]
\centering
\includegraphics[width=0.45\textwidth]{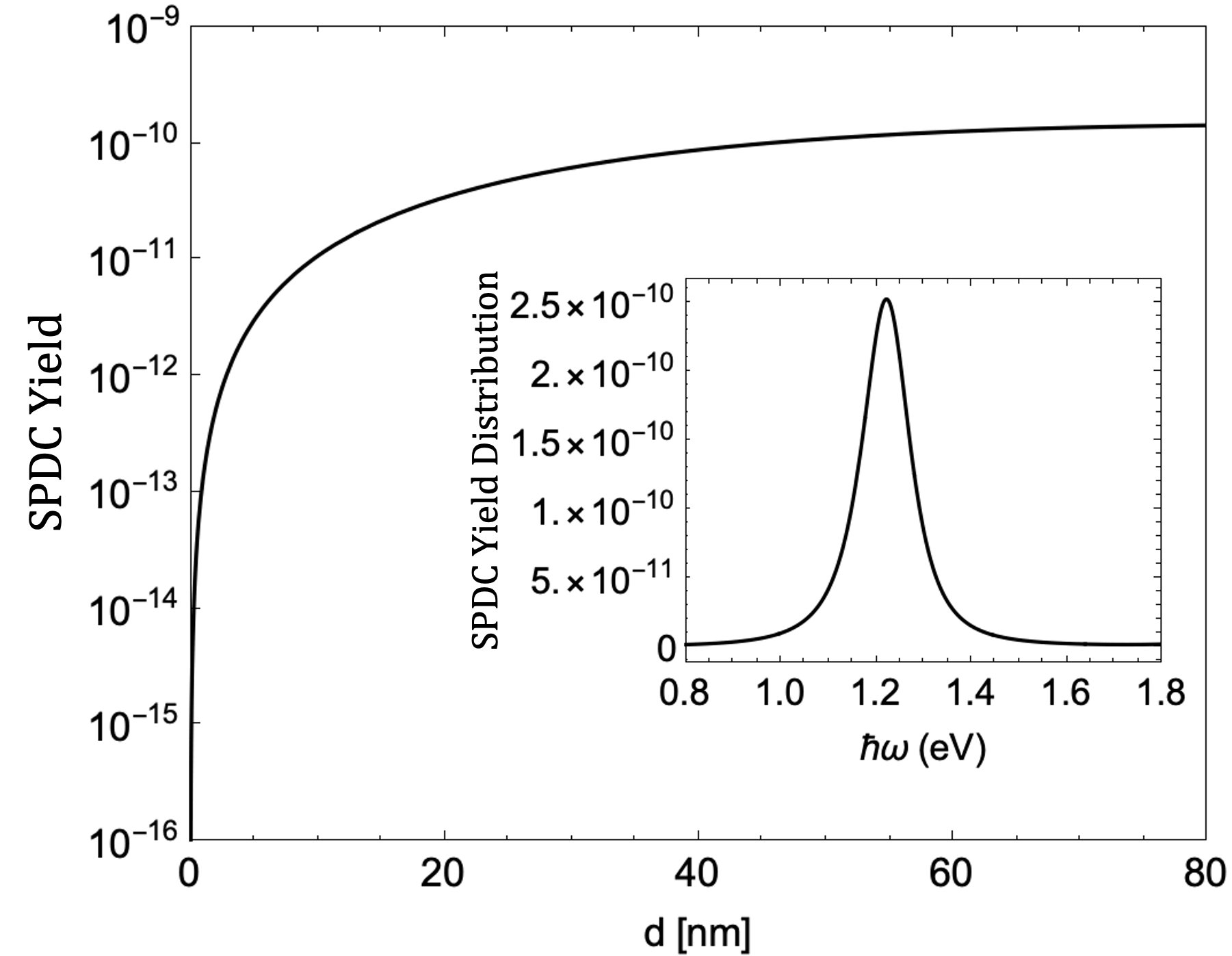}
\caption{The SPDC power conversion yield $\kappa(d)=I_{ti}/I_{is}$ calculated for Au~MNP. The inset shows the yield spectral distribution $d\kappa/d\omega$ peaked at the surface plasmon resonance.}
\label{fig:4}
\end{figure}

Ultimately, we investigate the SPDC yield of the Au MNP. In Sec.~5 of Supplement~1, we utilize the minimal model in a semiclassical approximation to derive the SPDC power conversion yield, defined as $\kappa(d)=I_{ti}/I_{is}$. Although this ratio shares similarities with the normalized idler intensity presented in Fig.~\ref{fig:2}(b), differences arise from the signal origin due to vacuum field fluctuations influenced by the plasmon resonance. Fig.~\ref{fig:4} presents the results of our calculations. According to the plot, the conversion efficiency saturates at $\sim 10^{-10}$ around $d\sim 80$nm. For comparison, a value $\kappa\sim 10^{-16}$ has been reported in studies of metal layers enhancing the extrinsic nonlinearity of a substrate and compared to $\kappa\sim 10^{-12}$ typical for 1~mm long BBO crystal~\cite{Loot_and_Hizhnyakov,Hizhnyakov_exp}. Our calculation shows the superior performance of $L$-shaped MNPs compared to the latter nanostructures for potential applications in probabilistic quantum light sources.

\section{Conclusion}

In conclusion, we have employed a minimal model to address DFG in an array of L-shaped MNPs demonstrating intrinsic second-order nonlinear response. Through the analysis of the spectral variation of the gain and loss terms near the local plasmon resonance, our calculations demonstrated that the gain behavior is primarily influenced by the wave vector mismatch leading to the gain regime above the fundamental frequency of the surface plasmon resonance. Notably, both our minimal model and the FDTD hydrodynamic-Maxwell model for Au MNPs demonstrate a tenfold increase in idler intensity over tens of nanometer scale. Our findings suggest that the parametric amplification regime becomes feasible on a scale of hundreds of nanometers while supported by high-intensity pumping. Specifically, achieving parametric amplification for a single layer of Au MNPs with a 30 nm thickness  requires the implementation of a photonic cavity or, alternatively, a multi-layered stack of arrays. Finlay, our calculations indicate that considered nanostructure holds promise for applications as a building block of quantum metasurfaces.

{\bf{Funding:}}  M.R.S. and M. C. were supported in part by the Air Force Office of Scientific Research under Grant No. FA9550-22-1-0175. Work by S.S. (M.R.C.) was supported (in part) by the Center of Nonlinear Studies via Laboratory Directed Research and Development program of Los Alamos National Laboratory under project number 3W400A-XXJ4. A.P. (M.R.C.) research was supported (in part) by the Laboratory Directed Research and Development program of Los Alamos National Laboratory under project number 20230363ER.

{\bf{Acknowledgments}} This work was performed, in part, at the Center for Integrated Nanotechnologies, an Office of Science User Facility operated for the U.S. Department of Energy (DOE) Office of Science by Los Alamos National Laboratory (Contract 89 233 218CNA000001) and Sandia National Laboratories (Contract DE-NA-0003525) (user project No. 2021BC0087).

{\bf{Data Availability:}} Data underlying the results presented in this paper are not publicly available at this time but may be obtained from the authors upon reasonable request.

{\bf{Disclosures:}} The authors declare no conflicts of interest.

\appendix

\section{Equations for electric field slowly varying amplitudes}

Let us define the pump, signal, and idler fields inside the metal nanoparticle (MNP) array as
\begin{eqnarray}
\label{Ep-SWA}
&~& E_p(\omega,z) = {\cal E}_{p}(z) e^{ik(\omega_p) z},
\\\label{Es-SWA}
&~& E_s(\omega,z) = {\cal E}_{s}(z) e^{ik(\omega) z},
\\\label{Ei-SWA}
&~& E_i(\omega_p-\omega,z) = {\cal E}_{i}(z) e^{ik(\omega_p-\omega)}. 
\end{eqnarray}
Here, ${\cal E}_{p}(z)$, ${\cal E}_{s}(z)$, and ${\cal E}_{i}(z)$ represent the slowly varying amplitudes for the pump, signal, and idler fields, respectively. Given $n(\omega)=\sqrt{1+\chi^{(1)}(\omega)}$ is the refractive index and $\chi^{(1)}(\omega)$ is defined in Eq.~(1) of the Letter, the quantity $k(\omega)=n\left(\omega\right)\frac{\omega}{c}$  is the $z$-component of the wave vector determining the propagation of the plane wave. In this notation, the spatial evolution of the signal and idler slowly varying amplitudes, assuming the non-depleted pump approximation, is governed by the following set of coupled equations~\cite{LoudonBook2000}
\begin{eqnarray}
\label{Es-eq}
\frac{d{\cal E}_s(z)}{dz} &=& g(\omega){\cal E}_i^*(z)e^{i\Delta k(\omega) z},
\\\label{Ei-eq}
\frac{d{{\cal E}_i^*(z)}}{dz} &=& g^*(\omega_p-\omega){\cal E}_s(z)e^{-i\Delta k(\omega) z}.
\end{eqnarray}
In Eqs.~(\ref{Es-eq}) and (\ref{Ei-eq}), the signal-idler coupling parameter is defined as 
\begin{eqnarray}
\label{gs-def}
 g(\omega) =i \frac{{\cal E}_{p}\omega}{n(\omega)c}  \chi^{(2)}(\omega_p;\omega;\omega_p-\omega),
\end{eqnarray}
and depends on the the second-order nonlinear susceptibility $\chi^{(2)}(\omega_p;\omega;\omega_p-\omega)$ defined in the Letter by Eq.~(6). The wave vector mismatch appearing in the equations above is
\begin{equation}
\centering
\label{dk-def}
\Delta k (\omega) = k(\omega_p) -k(\omega)-k^*({\omega_p-\omega}),
\end{equation}
where the pump is set within a lossless frequency range, ensuring that $k(\omega_p)$ has real values only. In contrast, the wave vectors of the signal, $k(\omega)$, and idler, $k(\omega_p - \omega)$, are complex since we tune them within proximity of the lossy plasmonic resonance.

Given the initial conditions ${\cal E}_s(0)= {\cal E}_{s0}$ and ${\cal E}_s(0)=0$, the solution of Eqs.~(\ref{Es-eq}) and (\ref{Ei-eq}) inside the array (0<z<d) is~\cite{LoudonBook2000}
\begin{eqnarray}
\label{sig-z}
{\cal E}_{s}(z) =& \\ \nonumber
&{\cal E}_{s0}\left[\cosh\left[\xi(\omega) z\right]
 -\frac{i \Delta k(\omega)}{2\xi(\omega)}\sinh\left[\xi(\omega) z\right]\right] e^{i\Delta k(\omega) z/2},
\\ \label{idler-z}
{\cal E}_{i}(z) =& {\cal E}_{s0}\frac{g(\omega_p-\omega)}{\xi(\omega)}
\sinh\left[\xi(\omega) z\right]~e^{i\Delta k(\omega) z/2},
\end{eqnarray}
respectively. Here, the shorthand notation $\xi(\omega)$ stays for the complex parameter defining gain and loss
\begin{eqnarray}
\label{xi-def}
\xi(\omega) = \sqrt{g(\omega)g^{*}(\omega_p-\omega)-\Delta k^2(\omega)/4}.
\end{eqnarray}

Introducing the following boundary conditions connecting the incident at $z=0$ pump ${\cal E}_{ip}$, signal ${\cal E}_{is}$, and transmitted at $z=d$ signal ${\cal E}_{ts}$ and idler ${\cal E}_{ti}$ amplitudes with the corresponding quantities inside the MNP array, 
\begin{eqnarray}
\label{Ep-BC0}
&~&{\cal E}_{p}= \frac{2{\cal E}_{ip}}{1+n(\omega_p)}, 
\\\label{Ep-BC0}
&~&{\cal E}_{s0}= \frac{2{\cal E}_{is}}{1+n(\omega)},
\\\label{Es-BCd}
&~&{\cal E}_{ts} = {\cal E}_{s}(d)\frac{2n(\omega)}{1+n(\omega)},
\\\label{Ei-BCd}
&~&{\cal E}_{ti} = {\cal E}_{i}(d)\frac{2n(\omega_p-\omega)}{1+n(\omega_p-\omega)},
\end{eqnarray}
respectively, and further substituting them into Eqs.~(\ref{sig-z}) and (\ref{idler-z}) we arrive at 
\begin{eqnarray}
\label{eq-sig-ratio}
&{\cal E}_{ts}(\omega) = \\ \nonumber
&\frac{4{\cal E}_{is} n(\omega)}{[1+n(\omega)]^2} \left(\cosh\left[\xi(\omega) d\right]
 -\frac{i \Delta k(\omega)}{2\xi(\omega)}\sinh\left[\xi(\omega) d\right]\right) e^{i\Delta k(\omega) d/2},
\\\label{eq-idler-ratio}
&{\cal E}_{ti}(\omega_p-\omega,d) = \\ \nonumber
&\frac{4{\cal E}_{is}n(\omega_p -\omega)g(\omega_p-\omega)}{\left[1+n^*(\omega)\right]\left[1+n(\omega_p -\omega)\right]\xi(\omega)}
\sinh\left[\xi(\omega) d\right]~e^{i\Delta k(\omega) d/2},
\end{eqnarray}
with
\begin{eqnarray}
\label{gs-mod}
 g(\omega) = {\cal E}_{ip} \frac{2i\omega\chi^{(2)}(\omega_p;\omega;\omega_p-\omega)}
        {c n(\omega)\left[1+n(\omega_p)\right]},
\end{eqnarray}
Ultimately, to obtain Eqs.~(2) -- (5) in the Letter, we convert the corresponding electric field amplitudes to the field intensities via the  formula $I=\varepsilon_o c |{\cal E}|^2/2$. 


\section{Model parameterization using FDTD simulations}

\begin{figure}[t!]
\centering
\includegraphics[width=0.45\textwidth]{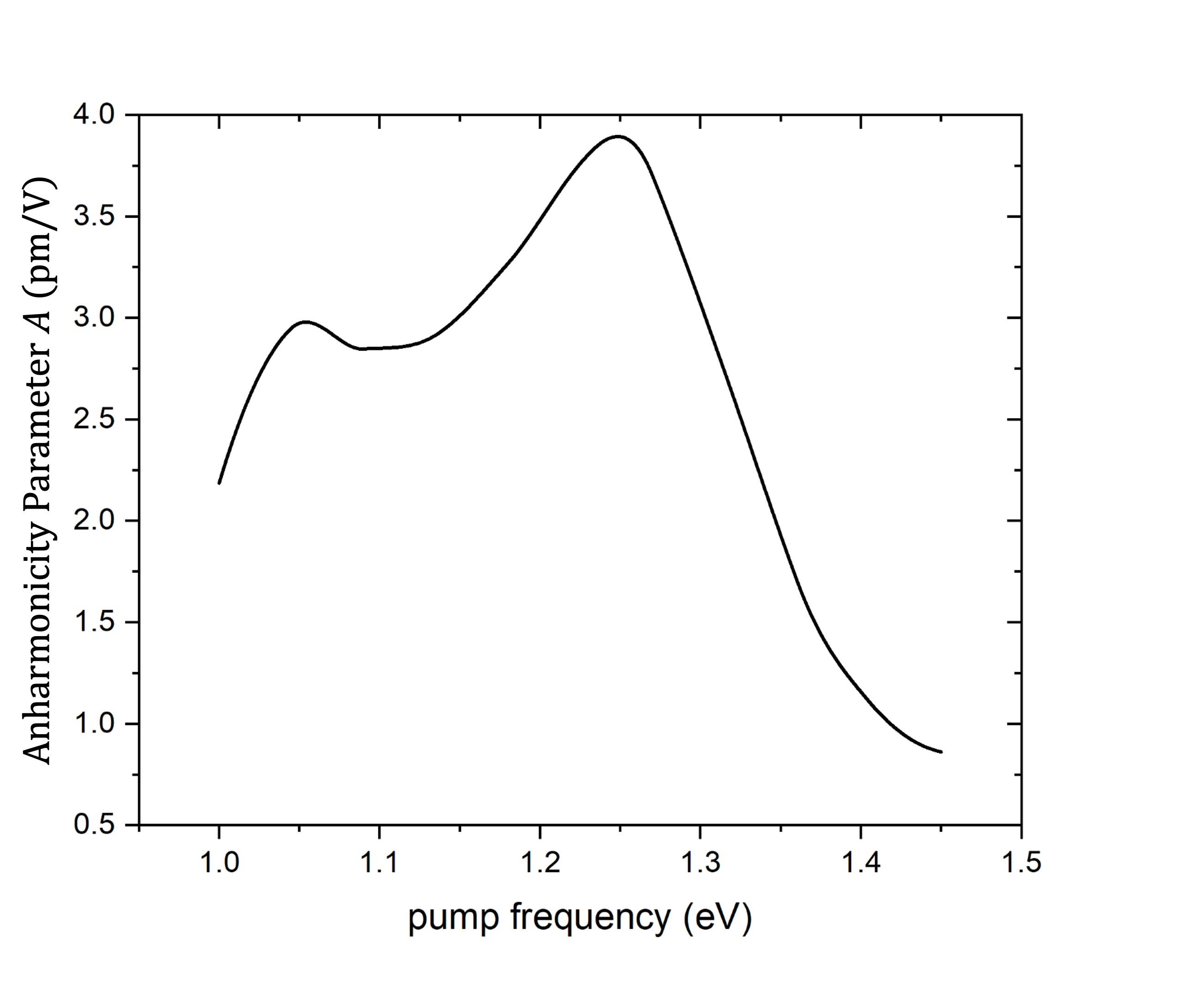}
\caption{The amplitude of the anharmonicity parameter $A$ extracted from second-harmonic simulations.}
\label{fig:S1}
\end{figure}

To parameterize our model, we adopted the methodology introduced in Ref~\cite{Minh:2023} and further developed in Refs.~\cite{SukharevJCP:2021,Sukharev_Drobnyh_Pachter_2022}. This approach involves utilizing FDTD calculations to determine the light scattering cross-section~\cite{Bohren_Book:2008} 
\begin{eqnarray}
\label{Cscat-def}
C_{\textrm{sca}}=\frac{1}{I_{\textrm{inc}}}\oint_A{\mathbf{S}_{\mathrm{sca}}}\cdot d\mathbf{A},
\end{eqnarray}
where $I_{\mathrm{inc}}$ represents the incident irradiance, and $\mathbf{S}_{\mathrm{sca}}$ is the Poynting vector of the scattered electromagnetic field integrated over a Gaussian surface, $A$. This surface is composed of two sections positioned in the far-field zone from the MNP array on both the input (detecting reflection) and output (detecting transmission) sides.

The MNPs can be regarded as a collection of point dipoles, with the linear response of each dipole described by polarizability $\alpha(\omega)$ and the second-order nonlinear response characterized by the first hyperpolarizability $\beta$. Subsequently, these quantities can be related to susceptibilities as
\begin{eqnarray}
\label{chi1-def}
\chi^{(1)} &=& \frac{\alpha}{2V},
\\\label{chi2-def}
\chi^{(2)} &=& \frac{\beta}{2V},
\end{eqnarray}
where $V$ is the MNP volume. The corresponding scattering cross-section~(\ref{Cscat-def}) for the first- and second-order responses have simple analytical representations
\begin{eqnarray}
\label{C-lin}
C_{\textrm{sca}}(2\omega) &=& \frac{\omega^4}{6\pi c^4}|\alpha(\omega)|^2,
\\\label{C-SHG}
C_{\textrm{sca}}(2\omega) &=& \frac{2\omega^4}{3\pi{}c^4}|\beta(2\omega)|^2 E_0^2.
\end{eqnarray}
 
We executed two sets of FDTD simulations using the hydrodynamic-Maxwell model (detailed in the Letter) to evaluate the total light scattering cross-section for the linear and second-harmonic responses of the Au NMP array. For the $y$-polarized surface-plasmon eigenmode of a single MNP, we fitted the linear scattering cross-section utilizing Eqs.~(\ref{C-lin}), (\ref{chi1-def}), and Eq.~(1) from the Letter. This fitting process yielded the plasmon energy $\hbar \omega_{sp} = 1.232$~eV, linewidth $\hbar \gamma_{sp}=0.077 $~eV, and transition dipole $p_{sp}= 4.75\times 10^{3}$~D. Subsequently, the second-harmonic response was fitted using Eqs.~(\ref{C-SHG}) and (\ref{chi2-def}). The results of this calculation are depicted in Fig~\ref{fig:S1}. According to the plot, the magnitude of $A$ reaches its maximum value of $\sim 4$ pm/V at the surface-plasmon resonance energy $\hbar\omega_{sp}=1.232$~eV. 

\section{Effect of Embedding Medium Refractive Index}

To account for the effect of a dielectric medium embedding the MNP array which is characterized by refractive index $n_m$, we replaced the linear susceptibility given by Eq.~(1) in the letter with
\begin{eqnarray}
\label{chi1-nd}
    \chi^{(1)}(\omega) = \frac{p_\texttt{sp}^2}{v\varepsilon_o n_m^2\hbar\left(\omega_\texttt{sp}-\omega -i\gamma_\texttt{sp}\right)},     
\end{eqnarray}
to modify the refractive index $n(\omega)=\sqrt{1+\chi^{(1)}(\omega)}$. Further incorporating the $n_m$ into the boundary conditions~(\ref{Ep-BC0})--(\ref{Ei-BCd}) we recast Eqs.~(2) and (3) provided in the Letter to the form 
\begin{align}
\label{Its-nm}
I_{ts} (\omega)&= I_{is}\left|\frac{4 n(\omega)}{[n_m+n(\omega)]^2}\right|^2\\ \nonumber
&\left|\cosh\left[\xi(\omega) d\right] \frac{i \Delta k(\omega)}{2\xi(\omega)}\sinh\left[\xi(\omega) d\right]\right|^2 e^{\textrm{Im}\Delta k(\omega) d},
\\\label{Iti-nm}
I_{ti} (\omega_p -\omega) &= I_{is}
    \left|\frac{4n(\omega_p -\omega)g(\omega_p-\omega)}{\left[n_m+n^*(\omega)\right] \left[n_m+n(\omega_p -\omega)\right]\xi(\omega)}\right|^2 \\ \nonumber
&\left|\sinh\left[\xi(\omega) d\right]\right|^2~e^{\textrm{Im}\Delta k(\omega)d},
\end{align}
where $\xi(\omega)$ defined in Eq.~(\ref{xi-def}) now contains the coupling parameter
\begin{eqnarray}
\label{gs-def}
 g(\omega) =\left(\frac{2I_{ip}}{\varepsilon_o c}\right)^{1/2} \frac{2i\omega\chi^{(2)}(\omega_p;\omega;\omega_p-\omega)}
        {cn(\omega)\left[n_m+n(\omega_p)\right]}.
\end{eqnarray}

\begin{figure}[t]
\centering
\includegraphics[width=0.45\textwidth]{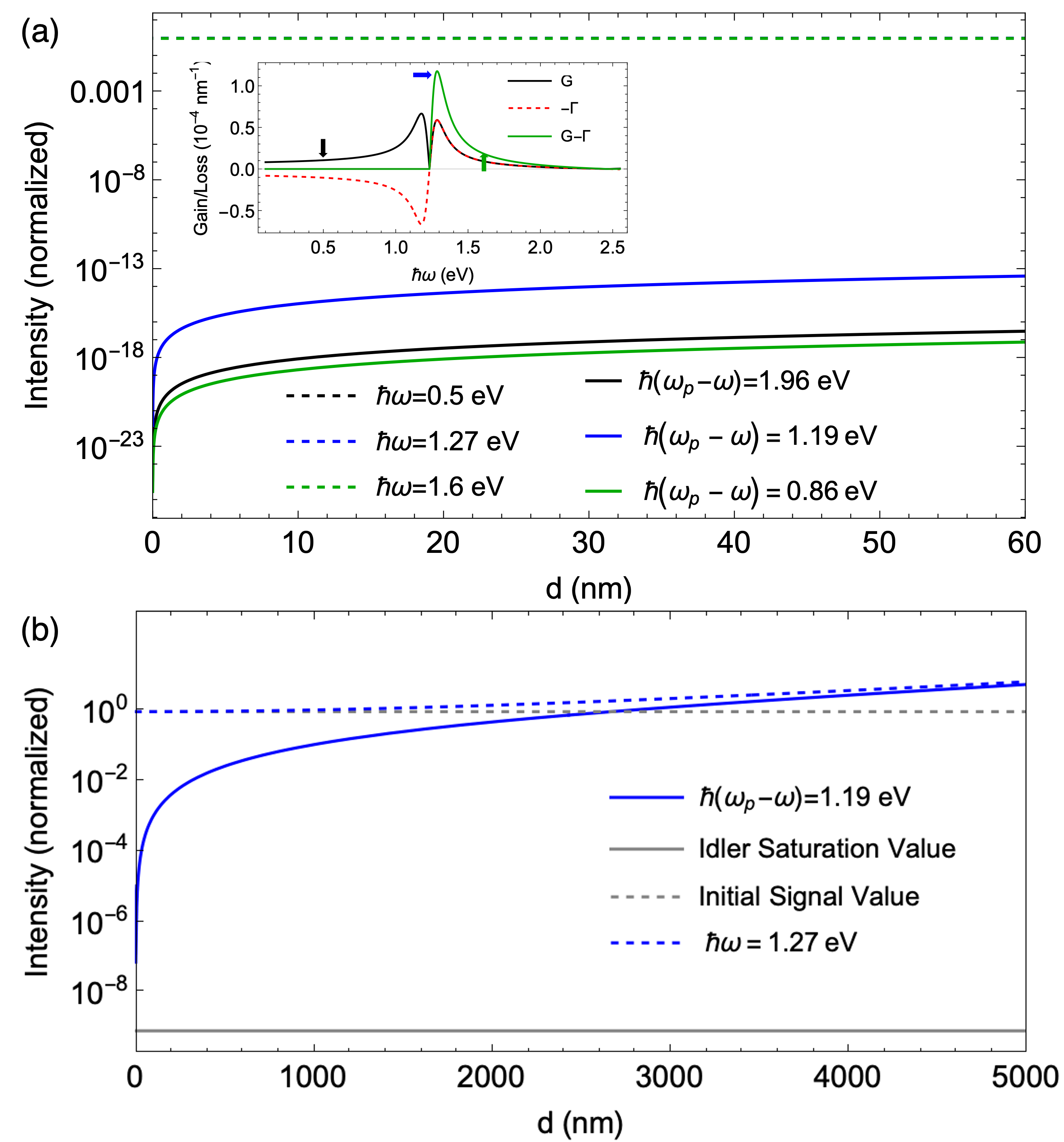}
\caption{(a) Signal and idler normalized intensities for the MNP lattice as discussed in the Letter, additionally embedded in a medium with a refractive index $n_m = 1.5$. The inset provides insight into the gain, loss, and net gain. (b) Similar to (a), signal and idler normalized intensities presented on an extended length scale, highlighting the parametric amplification threshold emerging around $d=3000$ nm. }
\label{fig:S2}
\end{figure}

In Fig.~\ref{fig:S2}, we plot the calculation outcomes using Eqs.~(\ref{Its-nm}) and (\ref{Iti-nm}) with all aforementioned modifications for the refractive index value $n_m=1.5$, which is a standard value for glass. The inset in panel (a) reveals a net gain value approximately four times lower than that depicted in Fig.~2(a) of the Letter for $n_m=1$. A further comparison of Fig.~\ref{fig:S2} with Fig.~2 in the Letter reveals a two-order of magnitude reduction for the idler at $\hbar (\omega_p-\omega) = 1.19$ eV with an array thickness of $d=60$ nm. Furthermore, it highlights the emergence of a signal amplification regime at the array thickness of $d\approx 10^3$ nm, in contrast to $d\approx 500$ nm for $n_m=1.0$.


\section{Power-conversion Efficiency}
\begin{figure}[t]
\centering
\includegraphics[width=0.45\textwidth]{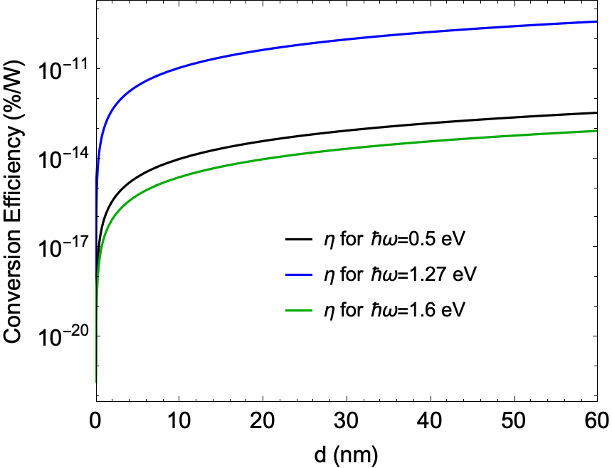}
\caption{The power conversion efficiency calculated for different signal energies $\hbar\omega$ based on the intensity ratios presented in Fig.~2(b) of the Letter and utilizing Eq.~(\ref{PCE-1}). }
\label{fig:S3}
\end{figure}

Defining the power-conversion efficiency for the difference-frequency generation process as the ratio 
\begin{eqnarray}
\label{PCE-def}
\eta(\omega_p-\omega) = \frac{P_{ti}(\omega_p-\omega)}{P_{is}(\omega)P_{ip}(\omega_p)}, 
\end{eqnarray}
where $P_{ti}(\omega_p-\omega)=A I_{ti}(\omega_p-\omega)$ represents the transmitted idler power normalized per incident signal $P_{is}(\omega)=A I_{is}(\omega)$ and pump $P_{ip}(\omega_p)=A I_{ip}(\omega_p)$ powers, with $A$ being the MNP cross-section area in the (x,y)-plane. Since we have already evaluated the ratio of transmitted idler and incident signal at a low conversion limit in the Letter's Fig.~2(b), the power conversion efficiency can be readily extracted from those calculations using the formula
\begin{eqnarray}
\label{PCE-1}
\eta(\omega_p-\omega) = \frac{I_{ti}(\omega_p-\omega)}{I_{is}(\omega)P_{ip}(\omega_p)}. 
\end{eqnarray}
Similar to the parameters adopted in the Letter, the electric field of the incident pump is set to ${\cal E}_{ip} = 1.25 \times 10^{-1}~\text{V/nm}$, and the MNP area to $A = 8.8 \times 10^3~\text{nm}^2$, corresponding to $P_{ip}=365~\textrm{mW}$. The results of these calculations are presented in Fig.~\ref{fig:S3}. 



\section{Semiclassical calculation of SPDC yield}

Assuming a constant amplitude, ${\cal E}_s$, for the signal over the distance $z$, we integrate Eq.~(\ref{Ei-eq}). The resulting electric field amplitude for the idler is 
\begin{align}
\label{Ei-lin}
{\cal E}_i^*(\omega_p-\omega, z) &=\\ \nonumber
&-\frac{{\cal E}^*_{p}(\omega_p)(\omega_p-\omega)}{n^*(\omega_p-\omega)c} 
 \chi^{(2)}(\omega_p;\omega,\omega_p-\omega){\cal E}_s(\omega)
\\\nonumber 
&\times \frac{z}{2}\text{sinc}\left(\frac{\Delta k(\omega)z}{2}\right)e^{-i\Delta k(\omega)z/2},
\end{align}
where $\text{sinc}(x)\equiv\sin(x)/x $. Using this result, we further introduce the idler auto-correlation function
\begin{align}
\label{Ei-cor-def}
&\left\langle{\cal E}_i^*(\omega_p-\omega, z){\cal E}_i(\omega'_p-\omega', z)\right\rangle = \\ \nonumber
&\frac{(\omega_p-\omega)(\omega_p'-\omega')}{n^*(\omega_p-\omega)n(\omega_p'-\omega')c^2}
    {\cal E}^*_{p}(\omega_p){\cal E}_{p}(\omega_p') \\\nonumber 
&\times    
    \chi^{(2)}(\omega_p;\omega;\omega_p-\omega)
    \chi^{(2)*}(\omega_p';\omega',\omega_p'-\omega') 
    \left\langle{\cal E}_s(\omega){\cal E}^*_s(\omega') \right\rangle \\\nonumber 
&\times
   \frac{z^2}{4}\text{sinc}\left(\frac{\Delta k(\omega)z}{2}\right) 
   \text{sinc}\left(\frac{\Delta k^*(\omega')z}{2}\right)
    e^{-i\left[\Delta k(\omega)-\Delta k^*(\omega')\right]z/2}.
\end{align}
The auto-correlation function of the signal appearing on the right-hand side of this expression, can be related with the imaginary part of the photon Green function, $G(z,z,\omega)$, incorporating the plasmonic response of the MNPs, via the fluctuation-dissipation  relation at zero temperature~\cite{NovotnyBook2012} 
\begin{eqnarray}
\label{Eop-def}
\left\langle{\cal E}{s}(\omega){\cal E}^{}{s}(\omega') \right\rangle =
\frac{\hbar\omega^2}{\pi\varepsilon_o c^2}\text{Im}\left[G(z,z;\omega)\right] \delta(\omega-\omega').
\end{eqnarray}

An observable quantity is the idler intensity which can be represented in terms of the idler field auto-correlation function  as~\cite{LoudonBook2000}
\begin{align}
\label{It-def}
I_i(z,t) &= 2\varepsilon_o c\int\int d\omega_p d\omega_p'\int\int d\omega d\omega'
\\\nonumber 
&\times \left\langle{\cal E}_i^*(\omega_p-\omega, z){\cal E}_i(\omega_p'-\omega', z)\right\rangle
\\\nonumber
&\times e^{i(\omega_p-\omega)t-ik^*(\omega_p-\omega)z}e^{-i(\omega_p'+\omega')t+ik(\omega_p'-\omega')z}.
\end{align}
By substituting Eq.~(\ref{Ei-cor-def}) into (\ref{Ei-lin}) and subsequently applying Eq.~(\ref{Eop-def}), we integrate over $d\omega'$ to attain the following expression
\begin{align}
\label{It-1}
I_i(z,t)&= \frac{\hbar z^2}{2\pi c^3}\int\int d\omega_p d\omega_p'e^{i(\omega_p-\omega_p')t}\int d\omega 
\\\nonumber &\times
    \frac{(\omega_p-\omega)(\omega_p'-\omega)}{n^*(\omega_p-\omega)n(\omega_p'-\omega)}
    {\cal E}^*_{p}(\omega_p){\cal E}_{p}(\omega_p') 
\\\nonumber &\times    
    \chi^{(2)}(\omega_p;\omega;\omega_p-\omega) 
    \chi^{(2)*}(\omega_p';\omega,\omega_p'-\omega)
\\\nonumber &\times
   \omega^2 \text{Im}\left[G(z,z;\omega)\right]
   \left |\text{sinc}\left(\frac{\Delta k(\omega)z}{2}\right)\right|^2 
\\\nonumber &\times    
    e^{-i\left[ k(\omega_p)- k(\omega_p')\right]z/2} e^{-\text{Im}[k(\omega)]z}   
\\\nonumber &\times
e^{-i\left[k^*(\omega_p-\omega)-k(\omega_p'-\omega)\right]z/2}.
\end{align}
This expression can be further averaged over a relatively large time interval by integrating both sides over time interval, leading to the emergence of the $2\pi\delta(\omega_p-\omega_p')$ function on the right-hand side. 

Dividing both sides of the equation by the integration interval yields the average idler intensity in the following functional form
\begin{eqnarray}
\label{It-2}
\frac{I_i(z)}{I_p} &=& \frac{\hbar z^2}{2\varepsilon_o c^4} \int\int d\omega_p d\omega~\omega^2
    \left|\frac{\omega_p-\omega}{n(\omega_p-\omega)}\right|^2
\\\nonumber&\times&
    w(\omega_p-\bar\omega_{p})
\\\nonumber &\times&    
    \left|\chi^{(2)}(\omega_p;\omega;\omega_p-\omega) \right|^2
\\\nonumber &\times&
    \text{Im}\left[G(z,z;\omega)\right]
   \left |\text{sinc}\left(\frac{\Delta k(\omega)z}{2}\right)\right|^2 
\\\nonumber &\times&    
    e^{-\text{Im}[k(\omega_p-\omega)+k(\omega)]z},
\end{eqnarray}
where, we explicitly introduce the Gaussian profile of the pump pulse using the envelope function  
\begin{eqnarray}
\label{w-def}
w(\omega_p-\bar\omega_p) = \frac{1}{\sqrt{2\pi}\sigma}e^{-\frac{\left(\omega_p-\bar\omega_p\right)^2}{2\sigma^2}}
\end{eqnarray}
with $\bar\omega_p$ representing the pulse central frequency, and $\sigma$ characterizing the pulse width. This envelope function is incorporated into Eq.~(\ref{It-2}) through the substitution $2\varepsilon_o c|{\cal E}_{p}(\omega_p)|^2=I_p w(\omega_p-\bar\omega_p)$, where $I_p$ denotes the mean pump intensity.  

Ultimately, under the condition of a narrow pulse width, the envelope function becomes $w(\omega_p-\bar\omega_p)=\delta(\omega_p-\bar\omega_p)$, leading to the simplification of Eq.~(\ref{It-2}) into the following form
\begin{eqnarray}
\label{It-3}
\frac{I_i(z)}{I_p}&=& \frac{\hbar z^2}{2\varepsilon_o c^4} \int d\omega~\omega^2
    \left|\frac{\omega_p-\omega}{n(\omega_p-\omega)}\right|^2
\\\nonumber&\times& 
    \left|\chi^{(2)}(\omega_p;\omega;\omega_p-\omega) \right|^2
    \text{Im}\left[G(z,z;\omega)\right]
\\\nonumber &\times&    
   \left |\text{sinc}\left(\frac{\Delta k(\omega)z}{2}\right)\right|^2 
    e^{-\text{Im}[k(\omega_p-\omega)+k(\omega)]z}.
\end{eqnarray}

Now, we introduce the SPDC power conversion yield for the MNP array of thickness $d$, as the ratio $\kappa(d)=I_{ti}/I_{ip}$, where $I_{ti}$ represents the idler intensity transmitted through the array and $I_{ip}$ is the incident pump intensity. Employing the boundary conditions~(\ref{Ep-BC0}) and (\ref{Ei-BCd}), we express the yield as
\begin{eqnarray}
\label{eta-def}
\kappa(d) = \left|\frac{\left[1+n(\omega_p-\omega)\right]n(\omega_p-\omega)}{1+n(\omega_p-\omega)}\right|^2\frac{I_i(d)}{I_p},
\end{eqnarray}
where either Eq.~(\ref{It-2}) or (\ref{It-3}) should be used to evaluate the ratio $I_i(z)/I_p$.

To determine a functional form for the Green function involved in Eqs.~(\ref{It-2}) and (\ref{It-3}) associated with the array of MNP, we introduce a basis set of modes 
\begin{eqnarray}
\label{umode-def}
\mathbf{u}_{\mathbf{k}\lambda}(\mathbf{r})=\mathbf{e}_{\mathbf{k}\lambda}\frac{e^{i\mathbf{k}\cdot\mathbf{r}}}{\sqrt{v}},
\end{eqnarray}
normalized per repeat unit volume $v$, where the polarization vector $\mathbf{e}_{\mathbf{k}\lambda}$ depends on $\mathbf{k}=\mathbf{k}(\omega)$ and $\lambda=1,2$. The modes satisfy the wave equation 
\begin{eqnarray}
\label{waveeq-umode}
\nabla\times\nabla\times \mathbf{u}_\mathbf{k}(\mathbf{r}) 
    - n^2(\omega_\mathbf{k})\frac{\omega_\mathbf{k}^2}{c^2}\mathbf{u}_\mathbf{k}(\mathbf{r})=0,
\end{eqnarray}
and orthogonality condition
\begin{eqnarray}
\label{ortho-mode-cond}
\int \mathbf{u}_\mathbf{k}(\mathbf{r})\cdot\mathbf{u}_\mathbf{k'}(\mathbf{r})d\mathbf{r}=\delta_{\mathbf{k}\mathbf{k'}}.
\end{eqnarray}
Furthermore, taking into consideration that the photon Green tensor satisfies the equation
\begin{align}
\label{waveeq-G}
&\nabla\times\nabla\times \mathbf{G}(\mathbf{r},\mathbf{r'};\omega) 
- n^2(\omega)\frac{\omega^2}{c^2}\mathbf{G}(\mathbf{r},\mathbf{r'};\omega)= \\ \nonumber
&\mathbf{I}\delta(\mathbf{r}-\mathbf{r'}),
\end{align}
where $\mathbf{I}=\text{diag}(1,1,1)$ and the refractive index $n(\omega) =\sqrt{1 + \chi^{(1)}(\omega)}$, we expand the Green tensor in the adopted mode basis, resulting in
\begin{eqnarray}
\label{G-gen}
\mathbf{G}(\mathbf{r},\mathbf{r'};\omega)=\frac{c^2}{v}\sum\limits_{\mathbf{k}\lambda} 
    \frac{\left(\mathbf{e}_{\mathbf{k}\lambda}\otimes\mathbf{e}_{\mathbf{k}\lambda}\right)
    e^{i\mathbf{k}\cdot(\mathbf{r}-\mathbf{r'})}}
    {n^2(\omega_\mathbf{k})\omega^2_\mathbf{k}-n^2(\omega)\omega^2}.
\end{eqnarray}

The Green tensor representation (\ref{G-gen}) encompasses full set of photon modes within the repeat unit volume. However, our down-conversion model only considers a narrow continuum of modes polarized in the $y$-direction and propagating along the $z$-axis with the wave vector $k(\omega)=n(\omega)\omega/c$. The associated component of the Green tensor can be easily extracted from Eq.~(\ref{G-gen}) by substituting the wave vector $\mathbf{k}\rightarrow k(\omega)$, coordinates $\mathbf{r}\rightarrow z$, and the integration measure $\sum_\mathbf{k}\rightarrow \sum_{k(\omega)}$. It is convenient to express the latter sum in continuous integral form as $(d/2\pi)\int_0^\infty (dk(\omega)/d\omega)d\omega$, where the prefactor $d$ represents the array thickness along the $z$-direction. Following these manipulations, the desired Green tensor component takes the following form
\begin{eqnarray}
\label{G-w}
G(z,z';\omega)=c^2\int\limits_0^\infty \rho(\omega') d\omega'
    \frac{e^{in(\omega')\omega'(z-z')/c}}
    {n^2(\omega')\omega'^2-n^2(\omega)\omega^2},
\end{eqnarray}
with the integration measure 
\begin{eqnarray}
\label{LDOS-rho}
\rho(\omega)=\frac{1}{2\pi c A}\left(n(\omega)+\frac{dn(\omega)}{d\omega}\omega\right),
\end{eqnarray}
accounting for the local density of states.

Finally, we present the explicit form of the local photon Green function at $z=z'$, 
\begin{eqnarray}
\label{G-w}
G(z,z;\omega)=c^2\int\limits_0^\infty
   \frac{\rho(\omega') d\omega'}{n^2(\omega')\omega'^2-n^2(\omega)\omega^2}.
\end{eqnarray}
where its imaginary part is employed to compute the SPDC yield using Eqs.~(\ref{eta-def}) along with Eqs.~(\ref{It-2}) and (\ref{It-3}).
The right-hand side integral in this expression is subject to numerical evaluation for the refractive index $n(\omega)$, whose parameterization is discussed in the Letter. In our calculations, we determined a single Au MNP yield. For this purposes, we set the unit cell volume $v$ in the susceptibility $\chi^{(1)}$ (Eq.~(1) in the Letter) to the MNP volume $V=2.64\times10^{5}~\text{nm}^3$. This is equivalent to considering a ``dense" array in which the pump absorption cross-section is due to the plasmonic NMPs only. For the finite pulse width $\sigma$, we scanned the range between 100 fs and 1.0 ns and found no significant variation in the SPDC yield compared to the $\delta$-function pulses. Therefore, only the latter case is discussed in the Letter.

\bibliography{main}

\end{document}